\documentstyle[twocolumn,prl,aps]{revtex}
\begin{document}
\title {The Effect of Spin-Splitting on the Metallic Behavior of a
Two-Dimensional System}
\draft
\author{S. J. Papadakis, E. P. De Poortere, H. C. Manoharan\cite{Hariaddress},
and M. Shayegan}
\address{Department of Electrical Engineering, Princeton University, Princeton,
New Jersey  08544}
\author{R. Winkler}
\address{Institut f\"ur Technische Physik III, Universit\"at
Erlangen-N\"urnberg, Staudtstr. 7, D-91058 Erlangen, Germany}
\date{\today}
\maketitle 
\begin{abstract}

Experiments on a constant-density two-dimensional hole system in a GaAs
quantum well reveal that the metallic behavior observed in the
zero-magnetic-field temperature dependence of the resistivity depends
on the symmetry of the confinement potential and the resulting spin-splitting of
the valence band.
\end{abstract}

For many years, it was widely accepted that there can be no metallic phase
in a disordered two-dimensional (2D) carrier system.  This was due to the
scaling arguments of Abrahams et al.\cite{Abrahamsscaling}, and the
support of subsequent experiments \cite{BishopTsui}.  In the last few years,
however, experiments on high quality 2D systems have provided us with reason to
re-visit the question of whether or not a metallic phase in a 2D system can
exist \cite
{KravMI,Popovic97,Coleridge97,Lam97,Hanein98,Simmons98,Papadakis98}.  Early
temperature dependence data by Kravchenko et al., from high-mobility silicon
metal-oxide-semiconductor field effect
transistors (MOSFETs), showed a drop in resistivity as the temperature
($T$) was reduced below 2 K.  This metallic behavior is the opposite of the
expected insulating behavior in which the resistivity should become infinite as
$T$ approaches zero.  In addition, the behavior not only was metallic
in a certain electron density range, but also scaled with a single parameter as
the density was reduced and the sample became insulating, suggesting a
true metal-insulator phase transition \cite{KravMI}.  

Since these experiments, the metallic behavior has been observed in
Si MOSFETS \cite{KravMI,Popovic97}, SiGe quantum wells \cite{Coleridge97,Lam97},
GaAs/AlGaAs heterostructures\cite{Hanein98,Simmons98}, and AlAs quantum wells
\cite{Papadakis98}, demonstrating that there are still some unsolved puzzles in
the fundamental nature of 2D carrier systems. 
Multiple mechanisms including electron-electron 
interaction \cite{Dobrosavljevic97}, spin-splitting \cite{Pudalov97}, and 
temperature dependence of traps \cite{Altshuler99} have been proposed as causes
for the metallic behavior, but no clear model has emerged which fully describes
this sizeable body of experimental data. The experiments reported here add to
our understanding by demonstrating a correlation between the
zero-magnetic-field spin-splitting and the metallic behavior.

Spin-splitting of carriers in a 2D
system at zero magnetic field is caused by the spin-orbit interaction and by an
inversion asymmetry of the potential in which the
carriers move \cite{Bychkov+}.  The energy bands are split into two
spin-subbands, which have different populations because
their energies at any non-zero $\vec{k}$ are slightly different.  The existence
of these spin-subbands has been well established both experimentally and
theoretically \cite{Bychkov+,Jusserand95,Dresselhaus92,spinsp,Lu98,Muraki98}.  

Our experiments are performed on high-mobility 2D hole systems in GaAs quantum
wells (QWs), chosen because they have a large inter-carrier separation $r_s$
\cite{rs}, have already shown metallic behavior \cite{Hanein98,Simmons98}, and
exhibit a large and tunable spin-splitting \cite{Lu98}.  In GaAs, the
spin-splitting arises from the inversion
asymmetries of the zincblende crystal structure and of the potential used to
confine the electrons to two dimensions.  The
asymmetry of the crystal structure is fixed, but the asymmetry of the confining
potential, and therefore the
spin-splitting, can be changed by applying an $\vec{E}$ perpendicular to the 2D
plane ($E_{\perp}$) using gates \cite{Lu98}.  Firstly we demonstrate that the
spin-splitting can be tuned while the density is kept constant, and then show
that the metallic behavior of the 2D holes is related to the amount of
spin-splitting observed.

Our samples are Si
modulation doped, 200 \AA-wide QWs grown by molecular beam epitaxy on the (311)A
surface of an undoped GaAs substrate.  QWs grown on the (311)
surface of GaAs exhibit a mobility
anisotropy believed to be caused by anisotropic interface roughness
scattering \cite{Heremans94}.  Photolithography is used to pattern L-shaped Hall
bars that allow simultaneous measurement of the resistivities along both the
high-mobility ($[\bar{2}33]$) and low-mobility ($[01\bar{1}]$) directions.   The
samples have metal front and back gates that control both the 2D hole density
and $E_{\perp}$.  Measurements are done in a dilution refrigerator at
temperatures from 0.7 K to 25 mK and in perpendicular magnetic fields ($B$) up
to 16 T.  We use the low-frequency lock-in technique, with a current of 10 nA,
to measure the longitudinal ($\rho$) and Hall resistivities.

To measure the spin-splitting we examine the low-$B$ $\rho$, or Shubnikov-de
Haas (SdH), oscillations \cite{spinsp,Lu98,Muraki98}.  The
frequencies of these oscillations, when multiplied by the level degeneracy
$e/h$, give the spin-subband densities, which are a
measure of the zero-$B$ spin-splitting.  To tune the splitting, we set the
front gate ($V_{fg}$) and back gate ($V_{bg}$) voltages, and measure
the resistivities as a function of $B$ on both arms of the Hall bar.
Then, at a small $B$, $V_{fg}$ is increased and the change in the hole
density noted.  $V_{bg}$ is then reduced to recover the original density.  This
procedure changes $E_{\perp}$ while maintaining the same density to within
1\%, and allows calculation of the change in $E_{\perp}$ from the way the
gates affect the density.  These steps
are repeated until we have probed the range of $V_{fg}$ and $V_{bg}$ that are
accessible without causing gate leakage.  This is done for two samples, from
different wafers, at 2D hole densities of  $2.3 \times 10^{11}$ cm$^{-2}$ ($r_s$
= 6.8) and $3.3 \times 10^{11}$ cm$^{-2}$ ($r_s$ = 5.7) \cite{rs}.  These
densities place the samples well into the metallic regime
\cite{Hanein98,Simmons98}.  The 25 mK mobilities of the two
samples are 83 and 51 m$^2$/Vs for the $[\bar{2}33]$ direction and
72 and 33 m$^2$/Vs for the $[01\bar{1}]$ direction respectively.  The results
from both samples are similar.  

Some of the low-$B$ $\rho$ data are shown (Fig. \ref{SdH}A) at a density of $3.3
\times 10^{11}$ cm$^{-2}$, from both the low- and high-mobility directions for
various sets of $V_{fg}$ and $V_{bg}$.  The top trace was taken with the sample
at a large positive $E_{\perp}$ (roughly 5,000 V/cm pointing towards the front
gate).  $E_{\perp}$ is reduced for the second trace, nearly zero in the middle
trace, and increasingly negative in the next two traces (it is about -6,000 V/cm
in the bottom trace).  The frequencies ($f_{SdH}$) of the
SdH oscillations are extracted by Fourier transforming the $\rho$ vs. $B^{-1}$
data in the range below 0.9 T.  The Fourier transforms of the low-mobility data,
(Fig. \ref{SdH}B), reveal how the spin-splitting changes as $E_{\perp}$
is changed from positive, through zero, to negative.  From the symmetry of
the data in Fig. \ref{splits}A, we estimate that $E_{\perp} = 0$ is near $V_{fg}
= -0.5$ V. 

The next part of the experiment involves measuring the $T$ dependence
of $\rho$ at $B = 0$ from 25
mK to about 0.7 K.  In the data from the low mobility direction (Fig.
\ref{SdH}C), the traces are separated
vertically and displayed next to the corresponding Fourier transforms in Fig.
\ref{SdH}B for clarity.  Their $\rho$ values at 25 mK and $B = 0$ ($\rho_0$) are
shown
on the $y$-axis.  The most striking feature is that the $T$ dependence of the $B
= 0$ resistivity is larger when the two $f_{SdH}$ peaks are well separated and
smaller when there is no separation.  It is clear that the magnitude of the $T$
dependence is correlated with that of the spin-splitting.  In order to
characterize this data in a simple way we plot $\Delta\rho^T/\rho_0$, the
fractional change in $\rho$ from 25 mK to 0.67 K (Fig. \ref{splits}B).  Figs.
\ref{splits}A and \ref{splits}B clearly show that $\rho$ increases more strongly
with $T$ as the difference between the SdH frequencies (${\Delta}f_{SdH}$) is
increased.  The same is true of the lower density sample.  Additionally,
$\Delta\rho^T/\rho_0$ is larger for lower density, consistent with previous
experiments on the metallic behavior
\cite{KravMI,Popovic97,Coleridge97,Lam97,Hanein98,Simmons98,Papadakis98}.

Finally, we attempt to more quantitatively determine what
the temperature dependence would be in the limit
of zero spin-splitting.  To this end, we have performed simulations
using self-consistent subband calculations that have no adjustable
parameters \cite{Lu98}.  They provide, for a given 2D hole density and
$E_{\perp}$, both simulated SdH oscillations and the zero-$B$
spin-subband population difference (${\Delta}p_s$).  Fourier transforms of the
simulated SdH oscillations show peak positions that agree very
well with those from the experimental data.
Furthermore, while the sum of the $f_{SdH}$ multiplied by $e/h$ gives the
total density, the calculations show that the ${\Delta}f_{SdH}$ underestimate
the zero-$B$ ${\Delta}p_s$ at all $E_{\perp}$.  In particular, the simulations
predict a finite zero-$B$ spin-splitting at $E_{\perp}$ = 0 (caused by the
inversion asymmetry of the GaAs crystal structure), while for $|E_{\perp}| <
1000$ V/cm the Fourier transforms of the simulated SdH oscillations and of the
experimental data show only one peak.  The measured
$\Delta\rho^T/\rho_0$ are plotted vs. the calculated ${\Delta}p_s$ (Fig.
\ref{DelRxx}) with the data for positive and negative $E_{\perp}$ 
separated on either side of the $x$-axis zero for clarity.  These results
suggest that $\Delta\rho^T/\rho_0$ would be zero or less than zero in the limit
of zero spin-splitting.

The data show clearly that the magnitude of $\Delta\rho^T/\rho_0$ increases
with the magnitude of $E_{\perp}$, which directly affects the amount of
spin-splitting.  Another parameter that is affected by $E_{\perp}$ is the
mobility.  However, as the $\rho_0$ values on the $y$-axis of Fig.
\ref{SdH}C show, the changes in mobility do not correlate with changes in
$\Delta\rho^T/\rho_0$.  For example, for three of the traces,
$\rho_0$ remains steady at 58.3 $\Omega$/sq., but $\Delta\rho^T/\rho_0$ changes.
Additionally $\Delta\rho^T/\rho_0$ is large when
the mobility is small, in contrast to the observation that the metallic behavior
becomes more pronounced as the mobility is improved \cite{Popovic97}.  Also, the
dependence of mobility on $E_{\perp}$ varies from sample to sample and
along different mobility directions within each sample \cite{mobility}.
Thus it is very unlikely that the mobility variation is causing the changes in
$\Delta\rho^T/\rho_0$. 

There are striking differences between the data along the low- and high-mobility
directions that add a new twist to this problem.  While the behaviors with
$E_{\perp}$ are qualitatively similar, $\Delta\rho^T/\rho_0$ is an order of
magnitude smaller in the high-mobility direction.  Additionally, the
low-mobility direction $\rho$ shows a strong positive magnetoresistance, not
present in the high-mobility $\rho$, that saturates at $\sim 0.1$ T.  Its
magnitude at 25 mK, ($\Delta\rho^B/\rho_0$) = [$\rho$(0.1 T) - $\rho$(0
T)]/$\rho_0$, is plotted (Fig. \ref{splits}C) and is remarkably similar
to $\Delta\rho^T/\rho_0$ in Fig. \ref{splits}B, despite the different origins
($B$ and $T$ dependence).  The difference in mobility for the two directions is
believed to come from an anisotropic interface roughness scattering that is due
to the interface morphology (see Ref. \cite{Heremans94} and references therein).
On the other hand, the Fourier transforms of the SdH oscillations
show that the spin-splitting is the same in both directions, so it is clear that
the changes in $\Delta\rho^B/\rho_0$ and $\Delta\rho^T/\rho_0$ are not
due to spin-splitting alone. It is possible that the directional differences in
the $B$- and $T$-dependencies of $\rho$ are due to an interplay between the
applied $B$-field or the spin-splitting and the anisotropic scattering mechanism
in the sample.  The precise nature of this interplay remains to be
discovered.  These observations point to the subtlety of the physics behind the
metallic behavior in this system.

We observe that the magnitude of the temperature dependence of 
the resistivity (the metallic behavior) increases with the magnitude of the
spin-splitting.  Previous experiments in applied $B$ fields have provided 
evidence that the spin 
of the carriers plays an important role in the metallic behavior.  However,
in these experiments the $B$ field, which increases
the spin-splitting, quenches the metallic 
state \cite{Simmons98,TiltB}.  
This inconsistency is surprising, and we surmise that in relation
to the metallic behavior
there is some fundamental difference between the zero-$B$ spin-splitting,
which is caused by the GaAs band structure and the confinment potential, and
the non-zero-$B$ spin-splitting, which is the difference in energy
between carriers with spins parallel and anti-parallel to the 
applied $B$ field \cite{splitexpl}.  Futhermore, our
results point out a complication in experiments that  
examined the dependence of the metallic behavior on the
carrier density:  they were changing the spin-splitting as they 
changed the density
\cite{KravMI,Popovic97,Coleridge97,Lam97,Hanein98,Simmons98,Papadakis98}.
Recognition of the fact that the zero-$B$ spin-splitting 
is on its own important will help untangle the causes of this unexpected
phenomenon.  Finally, we note that there are significant 
differences between the
low- and high-mobility direction data which have been
overlooked in previous experiments on GaAs 2D holes.  These
differences may also provide clues to  nature of the metallic behavior.

%

\begin{figure}
\caption{A:  Magnetoresistance traces, all at a density of $3.3 \times 10^{11}$
cm$^{-2}$, but at different values of $E_{\perp}$.  The data shown are from the
low-mobility $[01\bar{1}]$ (upper trace in each box) and high-mobility
$[\bar{2}33]$ directions.  B:  Fourier transforms of the Shubnikov-de Haas
oscillations, showing that the spin-splitting is being tuned through a
minimum.  C:  Temperature dependence of $\rho$ for the low-mobility
direction.  The traces are shifted vertically for clarity, with the value of
$\rho$ at 25 mK listed along the $y$-axis for each trace.  Each $\rho$
vs. $T$ trace is aligned with its corresponding Fourier transform.  B and C
together show that the magnitudes of the spin-splitting and the temperature
dependence are related.}
\label{SdH}
\end{figure}

\begin{figure}
\caption{A:  Measured Shubnikov-de Haas frequencies plotted as a function of
$V_{fg}$.  Each $V_{fg}$ has a corresponding $V_{bg}$ (top axis) that is used to
keep the density constant but vary $E_{\perp}$.  B:  $\Delta\rho^T/\rho_0$
vs. $V_{fg}$.  $\Delta\rho^T$ is the change in $\rho$ from $T = 25$
mK to 0.67 K and $\rho_0$ is $\rho$ at 25 mK and $B$ = 0 T.  C:
$\Delta\rho^B/\rho_0$ vs. $V_{fg}$.  $\Delta\rho^B$ is the
change in $\rho$ from $B = 0$ to 0.1 T at 25 mK.}
\label{splits}
\end{figure}

\begin{figure}
\caption{$\Delta\rho^T/\rho_0$ from the high density sample plotted vs. the
calculated zero-$B$ population difference of the spin-subbands.}
\label{DelRxx}
\end{figure}


\begin{thebibliography}{10}
\bibitem{mobility}
The behavior of the mobility with $E_{\perp}$ depends on the details of the
  disorder at the QW interfaces, and is therefore expected to be different from
  one sample to the next. Furthermore, one would expect, as we observe, a lower
  mobility when the QW is asymmetric because the carrier wave-function is
  pushed towards one of the interfaces, which are invariably somewhat
  disordered.
\bibitem{splitexpl}
The zero-$B$ spin-splitting is caused by spin-orbit interaction that is a
  complicated function of the carrier's momentum and the electrical potential
  it is traveling through. In contrast, the
  non-zero-$B$ spin-splitting (Zeeman splitting) is caused by an applied
  external field.
\end{thebibliography}
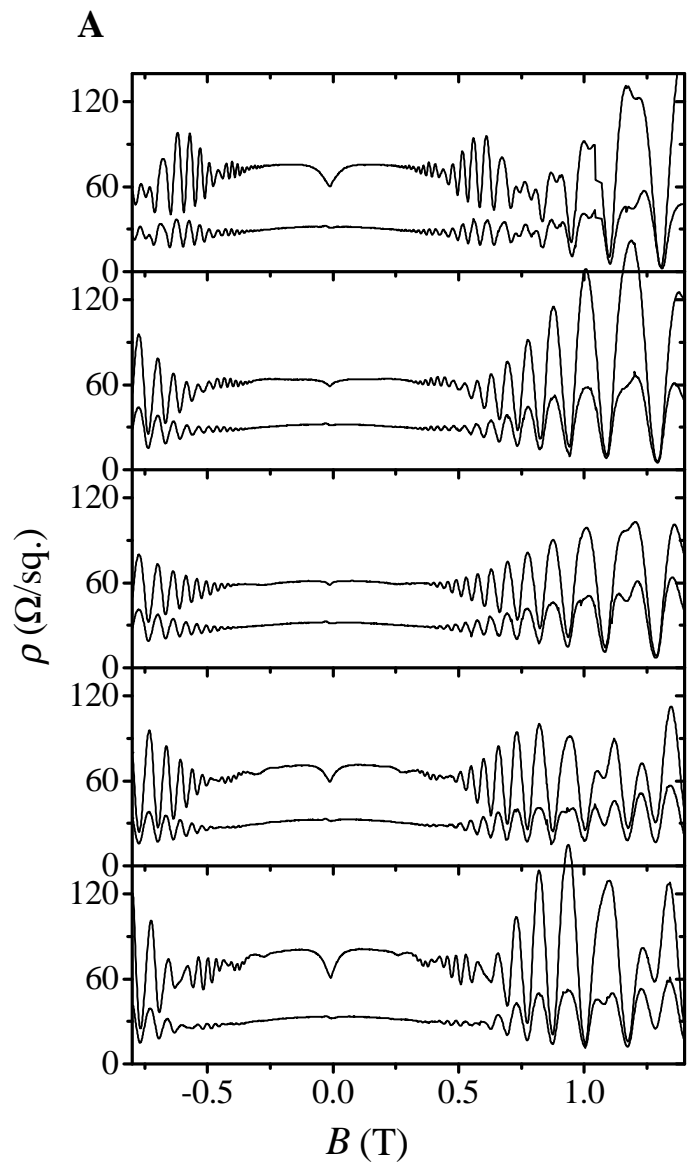
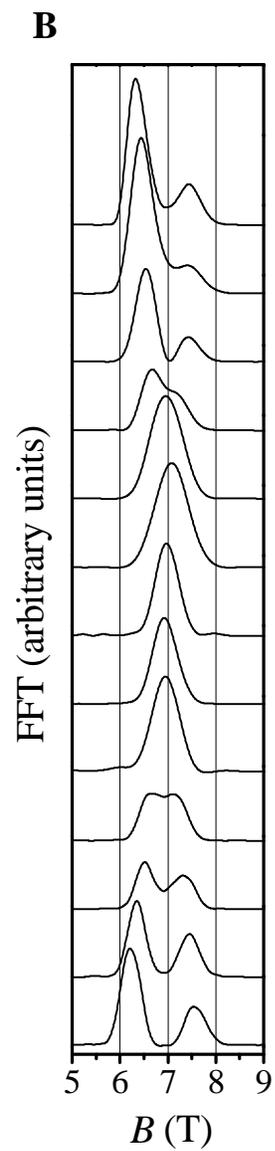
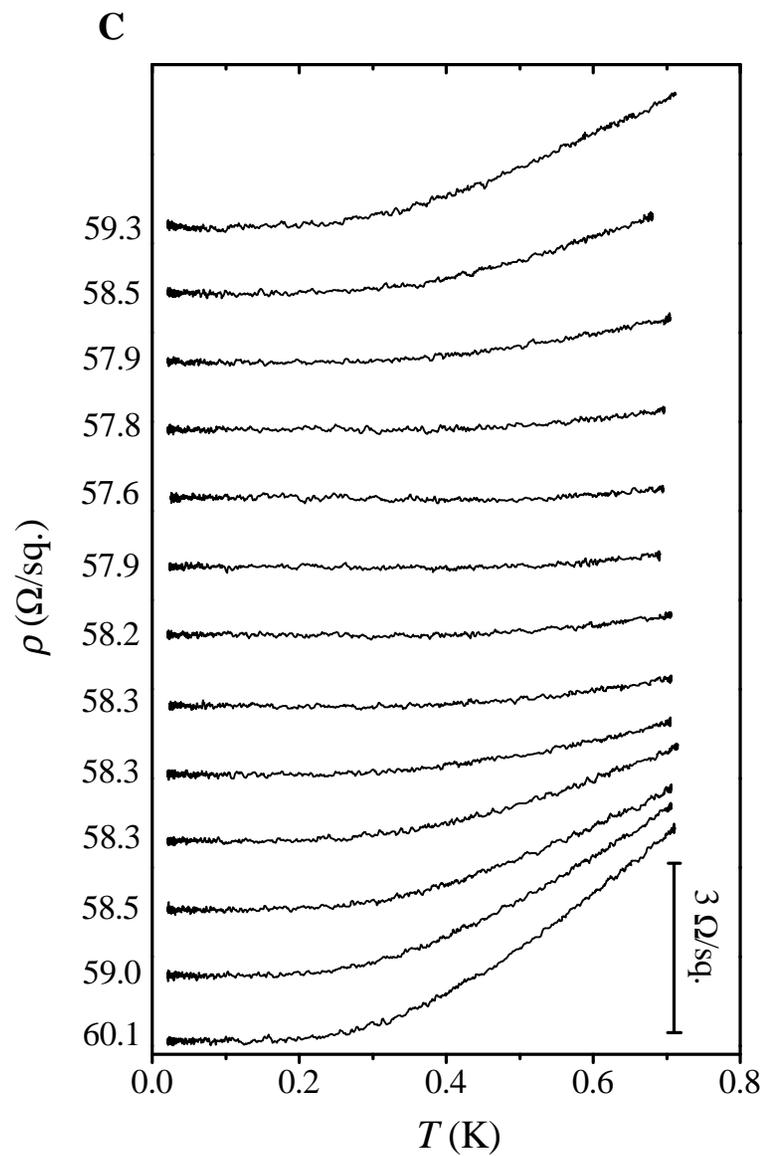
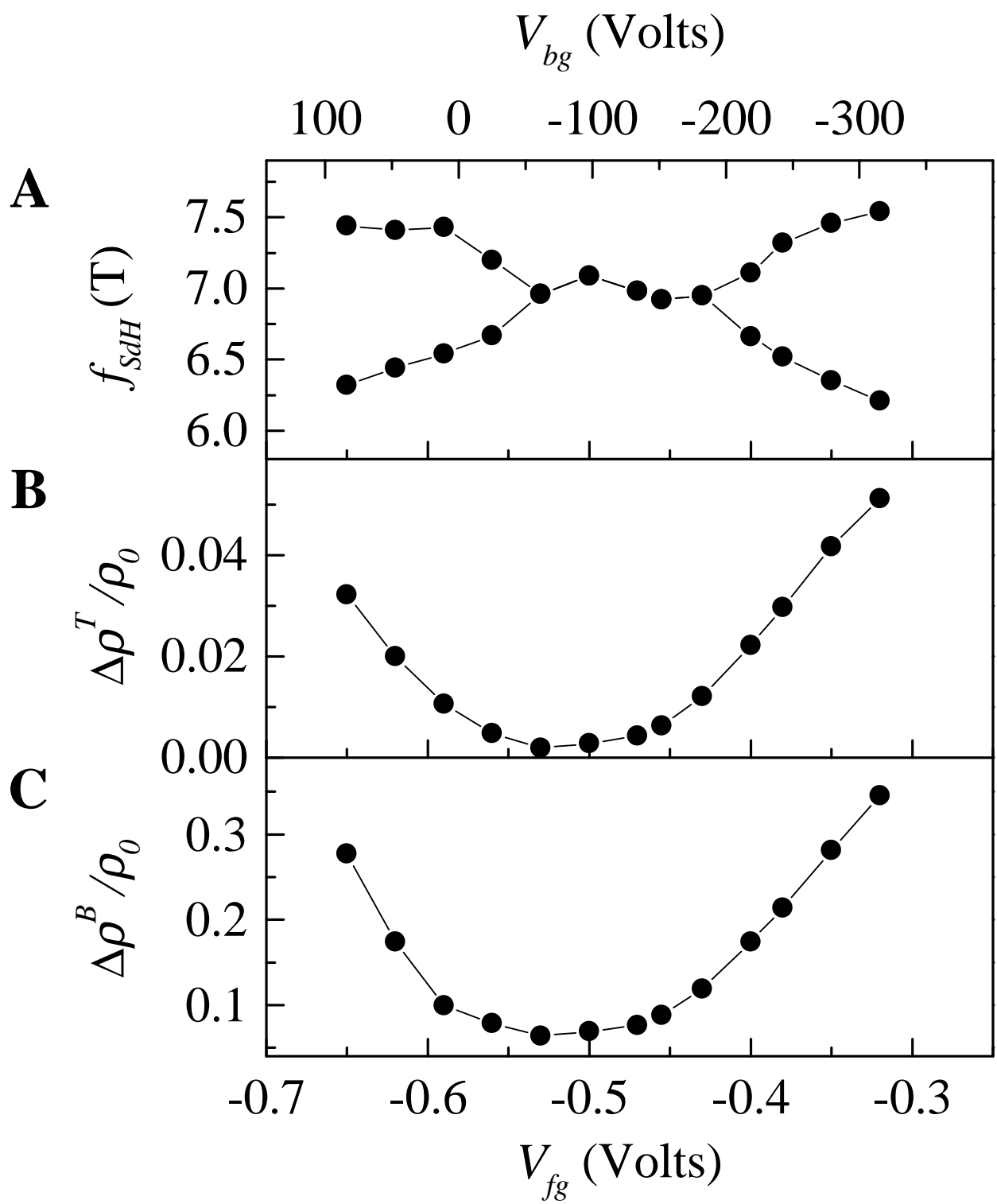
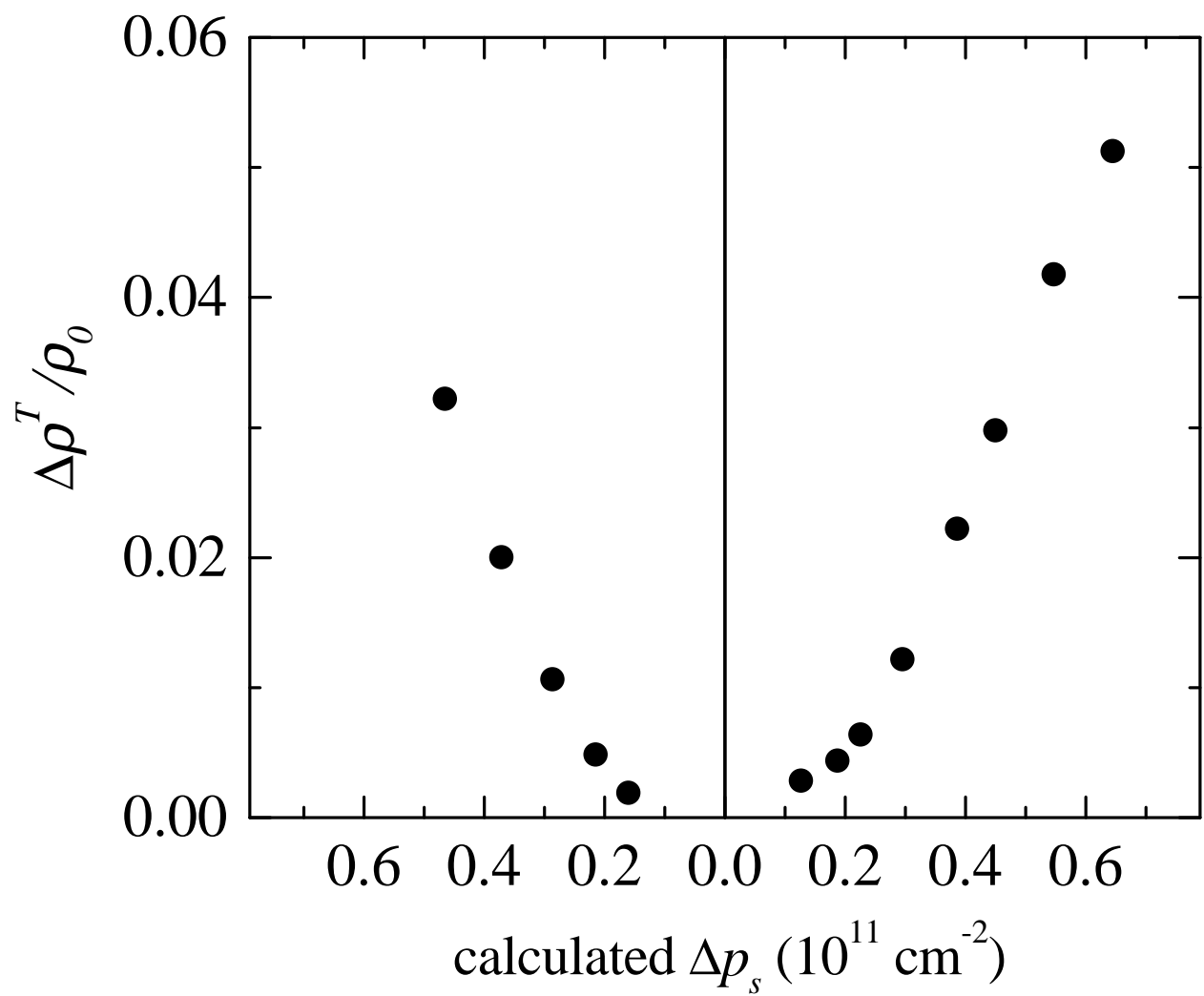
\end{document}